\documentstyle[11pt]{article}
\def\double{\baselineskip 24pt \lineskip 10pt} 

\def\be{\begin{equation}}
\def\ee{\end{equation}}
\def\bea{\begin{eqnarray}}
\def\eea{\end{eqnarray}}

\renewcommand{\theequation}{\arabic{section}.\arabic{equation}}

\newcommand{\w}{\omega}
\newcommand{\al}{\alpha}
\newcommand{\bp}{\beta_{+}}
\newcommand{\bm}{\beta_{-}}
\newcommand{\sig}{\sigma}
\newcommand{\th}{\theta}      
\newcommand{\ga}{\gamma}      
\newcommand{\half}{\frac{1}{2}}
\newcommand{\ka}{\kappa}
\newcommand{\nn}{\nonumber}
\newcommand{\co}{\rm{c}}
\newcommand{\si}{\rm{s}}

\def\case#1/#2{\textstyle\frac{#1}{#2}}
\def\k0{\kappa_{0}}

\begin{document}
\begin{titlepage}


\vspace{1in}

\begin{center}
\Large
{\bf Scale Factor Dualities in Anisotropic Cosmologies}

\vspace{1in}

\normalsize

\large{Dominic Clancy$^1$, James E. Lidsey$^2$ \& Reza Tavakol$^3$}

\normalsize
\vspace{.7in}

{\em Astronomy Unit, \\
School of Mathematical 
Sciences,  \\ 
Queen Mary \& Westfield College, \\
Mile End Road, \\
LONDON, E1 4NS, U.K.}

\end{center}

\vspace{1in}

\baselineskip=24pt
\begin{abstract}
\noindent
The concept of scale factor duality is considered within 
the context of the spatially homogeneous, vacuum 
Brans--Dicke cosmologies. 
In the Bianchi class A, it is found that 
duality symmetries exist for the types I, II, ${\rm VI}_0$, ${\rm VII}_0$, 
but not for types VIII and IX. The Kantowski--Sachs and 
locally rotationally symmetric Bianchi 
type III models also exhibit a scale factor duality, but   
no such symmetries are found for the Bianchi type V.
In this way anisotropy and spatial curvature may have
important effects on the nature of
such dualities.

\end{abstract}

PACS NUMBERS: 04.50.$+$h, 11.30.Ly, 95.30.Sf, 98.80.Hw

\vspace{.7in}
$^1$Electronic address: dominic@maths.qmw.ac.uk
 
$^2$Electronic address: jel@maths.qmw.ac.uk

$^3$Electronic address: reza@maths.qmw.ac.uk
\end{titlepage}

\double

\section{Introduction}

\setcounter{equation}{0}
 
\def\theequation{\thesection.\arabic{equation}}

The concept of {\em scale factor duality} 
has played a central role in string cosmology in recent years 
\cite{scalefactorduality,ts,gpr,meissner}. 
If one assumes a spatially isotropic and flat 
Friedmann--Robertson--Walker (FRW) universe,
the genus--zero effective superstring action is invariant under the 
discrete ${\rm Z}_2$ transformation \cite{scalefactorduality}
\be 
\label{1}
\bar{a} =a^{-1} , \qquad \bar{\Phi} = \Phi
- 6 \ln a 
\ee 
that inverts the cosmological 
scale factor, $a$, and shifts the value of the dilaton 
field, $\Phi$. Eq. (\ref{1})
maps an expanding cosmological solution onto a contracting 
one, and vice--versa. However, since 
the theory is also invariant under time reversal, $\bar{t}=-t$, 
the contracting 
solution may be mapped onto a new, expanding solution. 
Furthermore, the Hubble 
expansion parameter $H \equiv d \ln a /dt$ is invariant under the 
simultaneous application of scale factor duality and 
time reversal, but its first derivative changes sign. This implies that 
the new solution will represent a superinflationary 
cosmology characterized by the conditions $\dot{\bar{a}}>0, 
\ddot{\bar{a}}>0$ and 
$\dot{\bar{H}}>0$ if the original solution describes  a decelerated
expansion with $\dot{a}>0$ and $\ddot{a}<0$, and vice--versa, 
where a dot denotes differentiation with respect to cosmic time. 

It is this feature that forms the basis of the {\em pre--big bang 
scenario} \cite{scalefactorduality,pbb}. In this picture, the universe 
was initially in a state of low
curvature and weak coupling. The non--minimal coupling of the dilaton
field to the graviton, $g_{\mu\nu}$, 
drives an epoch of superinflation that represents 
the pre--big bang phase of the universe's history. 
The universe evolves into a 
regime of high curvature and strong--coupling
and to lowest--order in the string effective action, the end state
is a singularity in both the curvature and the coupling 
\cite{gracefulexit}. 
However, it is anticipated that such a singular state  
will be avoided when higher--order and quantum effects are included 
\cite{gmv,easther,lidsey}. 
The conjecture is that this will result 
in a smooth transition to the (duality related) 
post--big bang, decelerating solution. 

Since one of the main advantages of inflationary models in general is that 
they can explain the high degree of homogeneity, isotropy and 
spatial flatness in the observable universe, 
it is of importance to study the pre--big bang scenario in more 
general settings other than that of the spatially flat FRW cosmology. 
In the standard, 
chaotic inflationary scenario, 
such generalizations may have qualitatively 
significant consequences \cite{frag}.
Recently, Turner and Weinberg have studied the role
of spatial curvature within the context of the FRW universes 
\cite{tw}. 
They have argued that for 
certain initial conditions, spatial 
curvature may reduce the duration of the superinflationary 
expansion to such an extent that the horizon problem cannot 
be solved. Veneziano and 
collaborators, on the other hand, have considered 
the general effects of anisotropy and inhomogeneity on the scenario under the 
assumptions of small initial curvature and weak coupling 
and have shown that superinflationary 
expansion is possible in most regions of the universe \cite{ven97,vencol}. 

The purpose of this paper is to investigate whether 
the idea of scale factor duality may be extended 
to the spatially homogeneous and anisotropic Bianchi cosmologies 
\cite{esta,em,rs,mac}. 
This is interesting firstly because small anisotropies will 
inevitably be present in the real universe 
and secondly in view of the central role that scale factor duality
plays in the pre--big bang scenario. Bianchi 
cosmologies  represent the simplest deviations from the FRW environment
and they therefore allow us to gain insight into what may happen
as we deviate from the idealized FRW settings. 

Within the context of string theory, the scale factor duality,
Eq. (\ref{1}), is embedded within the symmetry group O(3,3) of the
anisotropic and spatially flat Bianchi type I background
\cite{ts,meissner}. More generally, there exists a target space
duality in the $\sigma$--model whenever there is an abelian symmetry
\cite{abelian}.  The group of duality transformations is O(d,d) when
$d$ abelian isometries are present. (For a recent review see, e.g., Ref. 
\cite{gpr}).  The procedure for establishing target space duality was
generalized to encompase non--abelian isometries in Ref. 
\cite{dq} and
considered further in Refs. \cite{grv,e,lozano,trace}. Gasperini, Ricci and
Veneziano discussed the spatially homogeneous cosmologies of arbitrary
dimension and, in particular, the dual to a Bianchi type V
cosmological solution \cite{grv}. However, they found that the dual model
did not satisfy the requirements for conformal invariance (vanishing
beta--functions) even when arbitrary shifts in the dilaton field were
allowed. Such a problem arises because the isometry group is
non--semisimple and this results in a trace anomaly that cannot be
absorbed by the dilaton \cite{trace,trace1}. The non--local nature of the
anomalous term in the Bianchi V solution was investigated by Elitzur
{\em et al.} \cite{e}. More general questions regarding the inverse
transformation \cite{79} and the quantum equivalence of the dual theories
have recently been addressed \cite{10} and non--abelian duals of the
Taub--NUT space have been derived in Refs. \cite{trace1,hew}.

The emphasis of the present paper is different in that we focus on the 
existence of dualities in the Brans--Dicke theory of gravity \cite{bd}. 
When the antisymmetric two--form potential vanishes,
the one--loop order beta--functions of 
string theory may be viewed 
as the field equations derived from the 
dilaton--graviton sector of the Brans--Dicke action, 
where the coupling constant between
the dilaton and graviton is given by $\omega =-1$ \cite{effective}. 
We consider all possible values of $\omega$, thereby placing 
the results from string cosmology into a wider context. Since 
we only consider the existence of scale factor dualities in this theory, 
we do not include the possible effects 
of the two--form potential. We also derive the conditions necessary for 
scale factor duality under the assumption that the dual and original 
spacetimes have the same isometry group. 
This differs from previous analyses in string theory, 
where in general the isometries of the dual background are different 
from those of the original. 

The Brans--Dicke theory is interesting for other reasons. 
It is the simplest example of a scalar--tensor theory
\cite{st} and represents a natural extension of Einstein gravity. 
The compactification 
of $(4+d)$--dimensional Einstein gravity on an isotropic, $d$-dimensional 
torus results in
a Brans--Dicke action, where the dilaton is related to the
radius of the internal space and $\omega =-1 +1/d$ \cite{higher}. 
Finally, the Brans--Dicke theory is consistent with the observational 
limits from primordial nucleosynthesis \cite{pn} and weak--field solar
system experiments \cite{solar} when $\omega > 500$ and it 
therefore represents a viable
theory of gravity in this region of parameter space. 


The spatially flat, $D$--dimensional FRW
Brans--Dicke cosmology exhibits a scale factor duality for all $\omega
\ne -D/(D-1)$ that reduces to Eq. (\ref{1}) when $\omega =-1$ and $D=4$
\cite{lidsey2}. Cadoni has recently studied the dualities of
$(2+1)$--dimensional Brans--Dicke gravity and found that there exists a
continuous O(2) symmetry in the theory when the metric corresponds to the 
three--dimensional equivalent of the Bianchi type I universe
\cite{cadoni}.  In this paper we find that
the Bianchi type I Brans--Dicke cosmology possesses
a continuous O(3) symmetry and show that this is
restricted to an O(2) symmetry for the type II model and to a discrete
duality for types ${\rm VI}_0$ and ${\rm
VII}_0$. The Kantowski--Sachs and locally rotationally symmetric (LRS)
Bianchi type III models also exhibit a scale factor
duality. On the other hand, no such symmetries exist for types V, VIII,
and IX. 

The paper is organised as follows. In section 2 we briefly summarize
the important features of the spatially homogeneous spacetimes and
derive the reduced Brans--Dicke actions for these models. In section 3,
we derive the symmetries of the types I, II, ${\rm VI}_0$ and ${\rm
VII}_0$ universes.  Types V, VIII and IX are investigated in section 4
together with the Taub universe, the Kantowski--Sachs model and the
LRS Bianchi type III cosmology.  We conclude in section 5.

\section{Spatially Homogeneous Brans--Dicke Cosmologies}
 
\setcounter{equation}{0}
 
\def\theequation{\thesection.\arabic{equation}}

The gravitational sector of the Brans--Dicke theory 
of gravity is \cite{bd}
\be
\label{bd}
S=\int d^4 x \sqrt{-g} e^{-\Phi} \left[ R -\omega \left( \nabla \Phi 
\right)^2 -2 \Lambda \right]\, ,
\ee
where $R$ is the Ricci curvature scalar of the spacetime with metric 
$g_{\mu\nu}$ and signature $(-, +, +, +)$, 
$g \equiv {\rm det} g_{\mu\nu}$, $\Phi$ is the dilaton field
and $\Lambda $ is the cosmological constant. 

Bianchi models are a class of 
spatially homogeneous cosmologies  
whose metrics admit a three--dimensional Lie 
group of isometries $G_3$ that
acts simply--transitively on 
three--dimensional space--like hypersurfaces of 
homogeneity \cite{esta,em,rs,mac}.
The line element for these 
cosmologies may be written as
\be
\label{lineelement}
ds^2 =-dt^2 + h_{ab} \omega^a \omega^b, \qquad a, b = 1,2,3\, ,
\ee
where $t$ represents cosmic time, 
$h_{ab}=h_{ab} (t)$ is the metric 
on the surfaces of homogeneity and $\omega^a$ are the corresponding
one--forms for each Bianchi type. 
If a three--form $\epsilon_{abc} =\epsilon_{[abc]}$ is specified on the Lie 
algebra of $G_3$, the antisymmetric structure constants
${C^a}_{bc} =-{C^a}_{cb}$ may be written as ${C^a}_{bc}=
m^{ad}\epsilon_{dbc} +{\delta^a}_{[b}a_{c]}$, where
$m^{ab}=m^{ba}$, $a_c \equiv {C^a}_{ac}$
and indices are raised and lowered with $h^{ab}$ and $h_{ab}$, respectively. 
The Jacobi identity ${C^a}_{b[c} {C^b}_{de]} =0$ then 
implies that $m^{ab}a_b =0$, 
i.e., that $a_b$ must be transverse to $m^{ab}$. 
The Lie algebra is in the Bianchi class A if $a_b =0$ and the 
class B otherwise \cite{em}.
A basis may be chosen without  loss of generality 
where $m^{ab}$ has a diagonal form with components $\pm 1$ or $0$
and $a_b =(A, 0, 0)$.  

For the Bianchi class A, the scalar 
curvature,  ${^{(3)}}R$,  of the three--surfaces is uniquely determined 
by the structure constants of the Lie algebra of $G_3$ \cite{wald}: 
\bea
{^{(3)}}R=-{C^a}_{ab}{{C^c}_c}^b + \frac{1}{2} {C^a}_{bc} {{C^c}_a}^b
-\frac{1}{4} C_{abc}C^{abc} \nonumber\\
\label{scalarcurvature}
= -h^{-1} \left( m_{ab}m^{ab} -\frac{1}{2}
m^2 \right)\,,
\eea
where $h^{-1} \equiv {\rm det}h^{ab}$. In the case of diagonal 
Bianchi models, the three--metric
may be parametrized in the form 
\be
\label{threemetric}
h_{ab}(t) = e^{2\alpha (t)} \left( e^{2\beta (t)} \right)_{ab}\,,
\ee
where
\be
\beta_{ab}={\rm diag} \left[ -2 \beta_+, \beta_+ -\sqrt{3} \beta_- , 
\beta_+ +\sqrt{3} \beta_- \right] 
\ee
is a traceless matrix that quantifies the anisotropy 
(shape change) of the models.
The `averaged' scale factor $a(t ) \equiv e^{\alpha (t) } = 
[ {\rm det} h_{ab} ]^{1/6}$
determines the behaviour of the effective 
spatial volume of the universe. The functions $a_i(t)$ defined 
by $h_{ab} \equiv {\rm diag} \left( a_1^2, a_2^2, a_3^2 \right)$ have the form 
\bea
\label{scalefactors}
a_1 = \exp[{\al-2\bp}]  \nonumber \\
a_2 = \exp[{\alpha + \bp -\sqrt{3}\bm}]  \nonumber \\
a_3 = \exp[{\alpha + \bp + \sqrt{3}\bm}] 
\eea
and may  be considered as the cosmological scale factors 
in the different
directions of anisotropy. 
The FRW cosmologies 
represent the isotropic limits of the Bianchi types I (and ${\rm VII}_0$) 
for $k=0$, type V for $k=-1$ and type IX for $k=+1$. 

The second--order field equations 
for  the  Bianchi class A  cosmologies 
may be consistently derived from a point Lagrangian  
by substituting the {\em ansatz} (\ref{lineelement}) 
into Eq. (\ref{bd}) and integrating over the spatial 
variables. This class includes types I, II, ${\rm VI}_0$, ${\rm VII}_0$, 
VIII and IX. Assuming that the dilaton is constant 
on the surfaces of homogeneity, we find that 
\be 
\label{bd1}
S= \int dt e^{3\alpha-\Phi}\left[ 
-6 \dot{\alpha}^2 +6\dot{\alpha}\dot{\Phi}+\omega\dot{\Phi}^2 
+6 \dot{\beta}_+^2+6 \dot{\beta}_-^2 
+{^{(3)}}R (\alpha,\beta_{\pm})-2\Lambda\right]\,,
\ee
where a boundary term has been ignored. 
The action for each Bianchi type A cosmology 
is therefore uniquely determined by the functional form of the 
three--curvature (\ref{scalarcurvature}). 

The Lagrangian formulation of the field equations for the class B models
is ambiguous because there exist spatial divergence
terms whose variations do not vanish, with the  
consequence that the variational principle does not
always result in  the correct field equations \cite{mac,sneddon,noether}. 
As a result, we will not discuss the class B models in detail. However,  
one class B model where a consistent Lagrangian 
can be derived is the Bianchi type V
with a metric of the form 
\be
\label{Vmetric}
ds^2=-dt^2 +e^{2\alpha (t)} \left[ dx^2 +e^{2x} \left( 
e^{-2\sqrt{3}\beta_-(t)} dy^2+ e^{2\sqrt{3}\beta_-(t)}dz^2 \right) 
\right] .
\ee
The Brans--Dicke action is formally given by Eq. (\ref{bd1}) with 
$\beta_+ =0$,
where the three--curvature 
is
\be
\label{curvature5}
{^{(3)}} R = - 6e^{-2\alpha}.
\ee

There also exists the Kantowski--Sachs model corresponding 
to the unique spatially homogeneous universe with no 
simply--transitive $G_3$ group of isometries \cite{ks}. The $G_3$ 
acts multiply--transitively on two--dimensional 
surfaces of maximal symmetry and positive curvature. 
The metric of the Kantowski--Sachs spacetime is 
\be
ds^2=-dt^2+a_1^2(t) dr^2+a_2^2(t)d\Omega^2_2   ,
\ee
where $d\Omega^2_2$ is the metric on the unit two--sphere. 
The LRS
Bianchi type III universe with ${m^a}_a =0$ may be viewed as 
the analytic continuation 
of the Kantowski--Sachs metric, where $d\Omega_2^2$ is replaced by the metric 
on a compact two--dimensional manifold of constant negative Ricci curvature 
$-2$. If we  express the two scale factors as
\bea
\label{kssf}
a_1 &\equiv& e^{\alpha -2 \beta} \nonumber \\
a_2 &\equiv& e^{\alpha +\beta},
\eea
the action (\ref{bd}) becomes
\be
\label{ksaction}
S = \int dt e^{3\alpha -\Phi} \left[ -6 \dot{\alpha}^2 +6 
\dot{\alpha}\dot{\Phi} + \omega \dot{\Phi}^2 +6 \dot{\beta}^2 + 
{^{(3)}}R (\alpha, \beta) -2\Lambda \right],
\ee
where the 
curvature potential is given by
\be
\label{kscurve}
{^{(3)}}R = 2k e^{-2\alpha -2\beta} 
\ee
and $k=+1,-1,0$ for the Kantowski--Sachs, LRS
Bianchi type III and axisymmetric Bianchi type I models, respectively.

\section{Symmetries of the Bianchi Class A Action}
\setcounter{equation}{0}
 
\def\theequation{\thesection.\arabic{equation}}

Throughout this paper 
we consider linear transformations on the variables appearing in 
Eqs. (\ref{bd1}) and (\ref{ksaction}). 
The symmetries of the action (\ref{bd1}) 
become manifest by rewriting it in terms of a {\em shifted scale factor} 
\be
\label{chi}
\chi \equiv \sqrt{\frac{3}{4+3\w}}\left [ \al+(1+\w)\Phi \right ]\,,
\ee
a {\em shifted dilaton field} 
\be
\label{sigma}
\sigma \equiv \kappa^{-1} \left ( \Phi - 3 \alpha\right ) , \qquad 
\kappa \equiv \sqrt{\frac{4+3\omega}{3+2\omega}}
\ee
and the rescaled anisotropy parameters 
\be
\label{beta12}
\beta_1 \equiv \sqrt{6} \beta_+ , \qquad  
\beta_2 \equiv \sqrt{6} \beta_-.
\ee  
Throughout we shall take $\omega > -3/2$
and exclude the particular value of $\omega = -4/3$.
The averaged scale factor and dilaton are given by
\bea
\label{average}
\alpha &=& -\frac{(1+\omega)}{\sqrt{(3+2\omega )(4+3\omega )}} 
\sigma +\frac{1}{\sqrt{3(4+3\omega )}} \chi  \\
\label{dilaton}
\Phi &=& \frac{1}{\sqrt{(3+2\omega )(4+3\omega )}} \sigma +
\sqrt{\frac{3}{4+3\omega}} \chi
\eea
and action (\ref{bd1}) then takes the form
\be
\label{bd2}
S=\int dt e^{-\kappa \sigma} \left[ -\dot{\sigma}^2 
+ \dot{\chi}^2 + \dot{\beta}^2_1 + \dot{\beta}^2_2 +{^{(3)}}R(\chi , 
\sigma , \beta_i ) -2 \Lambda \right]. 
\ee
We remark that the theory is invariant under time reversal, $\bar{t}=-t$. 

The action for the FRW cosmologies is given by Eq.(\ref{bd2}), where $\beta_i=0$ and ${^{(3)}}R=6ke^{-2\alpha}$. The kinetic sector of the action is then invariant under a discrete ${\rm Z}_2$ 
scale factor duality~\cite{scalefactorduality,lidsey2}
\be
\label{sfdfrw}
\bar{\chi} =-\chi , \qquad \bar{\sigma} = \sigma.
\ee
Eq. (\ref{average}) implies that the three--curvature is not invariant 
under (\ref{sfdfrw}) and the duality is therefore broken in the spatially 
curved isotropic universes. We now proceed to investigate  the symmetries 
of the Bianchi class A cosmologies.
\subsection{Bianchi Type I}
When the cosmological 
constant is non--zero, any symmetry of the theory that 
leads to non--trivial transformations of the dynamical 
degrees of freedom 
must leave the shifted dilaton field 
invariant, i.e., $\bar{\sigma} =\sigma$. This 
implies that the symmetries involve 
the remaining three variables, 
$\{ \chi, \beta_i \}$. 

The eigenvalues of $m^{ab}$ are all zero for the Bianchi  type I 
and the three--curvature therefore vanishes. It follows from 
Eq. (\ref{bd2}) that the action is symmetric under a 
continuous $\rm{O(3)}$ 
transformation. 
The standard realization for the action of this group on the variables 
$\{\chi,\beta_i\}$ is given by  
\be
\label{standard}
\bar{V} = R V, \qquad V \equiv 
{\left(
   \begin{array}{c}
      {\beta}_2 \\ {\beta}_1 \\ {\chi}
   \end{array} \right)}   ,
\ee
where 
\be 
\label{SO3}
R \equiv \pm 
{\left( 
  \begin{array}{ccc}
	-\si(\ga)\si(\psi)+\co(\ga)\co(\psi)\co(\theta)  &
	-\co(\ga)\co(\theta) \si(\psi)-\si(\ga)\co(\psi) & 
\co(\ga)\si(\theta) \\ 
	-\co(\ga)\si(\psi)+\co(\psi)\si(\ga)\co(\theta)  &
	-\si(\ga)\co(\theta) \si(\psi)-\co(\ga)\co(\psi) & \si(\ga)
\si(\theta) \\ 
	-\si(\theta) \co(\psi) & \si(\theta) \si(\psi) & \co(\theta)  
  \end{array}
\right)} ,
\ee
$\{\th,\ga,\psi\}$ are the three Euler angles, ${\rm s} 
(x)\equiv\sin x$ and 
${\rm c}(x)\equiv \cos x$. 
The group $\rm{O(3)}$ has two connected components
comprising the
subgroup $\rm{SO(3)}$, which is isomorphic to the group of proper rotations in 
three dimensions, and the  {\it set} of reflections. 
It is isomorphic to
${\rm SO(3)} \times {\rm G}_2$, where
${\rm G}_2$ is the matrix group of 
order 2 with elements $\{1_3$, $-1_3\}$ and $1_3$ is the $3\times 3$
identity matrix. Hence,  the proper 
rotations of $\rm{O(3)}$ are given by the positive
sign in 
Eq.  (\ref{SO3}) and the reflections are given by the 
negative sign. 

Substituting Eqs. (\ref{standard}) 
and (\ref{SO3}) into Eqs. (\ref{beta12})--(\ref{dilaton})
implies that the averaged scale factor, dilaton
and anisotropy parameters transform to 
\bea
\label{baralpha} 
\bar{\al} =
\frac{3(1+\w)\pm\cos\th}{4+3\w} \alpha \pm
\frac{\sqrt{8+6\w}\sin\th\sin\psi}{4+3\w} \beta_+ 
\mp \frac{\sqrt{8+6\w}\sin\th\cos\psi}{4+3\w} \beta_-\nonumber \\
\pm \frac{(1+\w)(\cos\th \mp 1)}{4+3\w} \Phi \\ 
\label{barphi}
\bar{\Phi} =
\pm \frac{3(\cos\th\mp 1)}{4+3\w} \alpha \pm
\frac{3\sqrt{8+6\w}\,\sin\th\sin\psi}{4+3\w} \beta_+\mp 
\frac{3\sqrt{8+6\w}\,\sin\th\cos\psi}{4+3\w} \beta_- \nonumber \\ 
\pm\frac{(3(1+\w)\cos\th\pm 1)}{4+3\w} \Phi  \\ 
\label{barplus}
\bar{\beta}_+ =
\pm \frac{\sqrt{8+6\w}\sin\ga\sin\th}{2(4+3\w)} \al \pm
(\cos\ga \cos\psi-\sin\ga\cos\th\sin\psi) \bp \nonumber \\ 
\pm (\cos\ga\sin\psi +\cos\psi\sin\ga\cos\th) \bm \pm
\frac{\sqrt{8+6\w} (1+\omega ) \sin\ga\sin\th}{2(4+3\w)} \Phi \\
\label{barminus}
\bar{\beta}_- =
\pm \frac{\sqrt{8+6\w}\cos\ga\sin\th}{2(4+3\w)}\al
\mp(\cos\ga\cos\th\sin\psi + \sin\ga\cos\psi)\bp \nonumber \\
\pm (\cos\ga\cos\psi\cos\th - \sin\ga \sin\psi) \bm 
\pm\frac{\sqrt{8+6\w}(1+\omega )\cos\ga\sin\th}{2(4+3\w)} \Phi ,
\eea
where the upper (lower) signs correspond to the rotations (reflections) 
of the $\rm{O(3)}$ symmetry. The corresponding 
transformations for the scale factors may then be derived by 
substituting Eqs. (\ref{baralpha})--(\ref{barminus}) into Eq. 
(\ref{scalefactors}). 

The effect of introducing anisotropy in this fashion is to 
extend the discrete ${\rm Z}_2$ duality (\ref{sfdfrw}) of the $k=0$ 
FRW model to a continuous $\rm{O(3)}$ symmetry. The duality (\ref{sfdfrw})
may now be interpreted as the subgroup of $\rm{O(3)}$ 
that leaves the anisotropy parameters at zero and inverts 
$e^{\chi}$. For example, one may specify the Euler angles to be 
$\theta= 2n\pi$ and 
$\gamma=  \psi=(-1)^n(2n+1)\pi/2$, where $n$ is a positive integer,
and consider the lower sign in Eq. (\ref{SO3}), 
since this implies that $R={\rm diag} (1,1,-1)$. When $R=-1_3$
and $\omega=-1$, we recover the analogue of the
scale factor duality (1.1) for the 
Bianchi type I string cosmology,  where all three scale factors are 
simultaneously inverted, $\bar{a}_i = a^{-1}_i$ \cite{ts,meissner}.

The kinetic sectors of all Brans--Dicke Bianchi class A actions 
are invariant under 
the transformations (\ref{baralpha})--(\ref{barminus}). 
However, the curvature 
potential (\ref{scalarcurvature}) is not in 
general symmetric under the full 
group of transformations. Depending on the Bianchi type, this 
imposes further restrictions on the symmetries of the model.

\subsection{Bianchi Type II}
We now consider the Bianchi II universe, where $m^{ab} = 
{\rm diag} ( 1, 0, 0)$ and the curvature potential is
\be
\label{curvature2}
{^{(3)}} R =-\half\exp[-2\al -8\beta_+ ].
\ee
Since the potential is independent of $\beta_-$, the action is symmetric 
under the discrete ${\rm Z}_2$ symmetry $\bar{\beta}_- = - \beta_-$, as in 
the general relativistic case. This corresponds to the simultaneous 
interchange of the cosmological scale factors $a_2 \leftrightarrow 
a_3$. Applying the change of variables defined in Eqs. 
(\ref{beta12})--(\ref{dilaton}) implies that
\be
\label{IIpotential}
{^{(3)}}R(\chi,\sigma,\beta_1 )=
-\half\exp\left[ C_1\sigma-\frac{8}{\sqrt{6}}
\left(C_2\chi+\beta_1\right)\right]\,,
\ee
where 
\bea
\label{c1}
C_1 &\equiv& \frac{2(1+\omega )}{\sqrt{(3+2 \omega )(4+3\omega )}} \\
C_2 &\equiv& \frac{1}{2\sqrt{8+6\omega}}.
\eea
The non--trivial curvature potential (\ref{IIpotential})
imposes an extra constraint 
on the group of transformations that leaves the action (\ref{bd2}) 
invariant. 
As well as requiring the kinetic sector to be invariant, 
we also require the constraint 
\be
\label{IIconstraint}
C_2 \bar{\chi} +\bar{\beta}_1 = C_2 \chi + \beta_1
\ee
to be satisfied. 

A symmetry of the theory may be found by 
performing  the field redefinitions
\bea
\label{xII}
x &\equiv& C_2 \chi + \beta_1  \\
\label{yII}
y &\equiv& \frac{2}{\sqrt{3}} \left( \frac{8+6\omega}{11+8\omega} 
\right)^{1/2} ( \chi -C_2 \beta_1 ).
\eea
Action (\ref{bd2}) transforms to
\be
\label{act2}
S =\int dt e^{-\kappa \sigma} \left[ 
\frac{4}{3} \left( \frac{8+6\omega}{11+8\omega} \right) \dot{x}^2 
+\dot{y}^2 + \dot{\beta}^2_2 -\dot{\sigma}^2 -\frac{1}{2} 
e^{C_1 \sigma 
-8x/\sqrt{6}} -2 \Lambda \right].
\label{S[xy]}
\ee
Thus, Eq. (\ref{act2}) is invariant 
under a continuous 
$\rm{O(2)}$ transformation that acts non--trivially 
on the variables $\{y,\beta_2\}$ and leaves 
 $\sigma$ and $x$ invariant. We may express the symmetry group 
as $\rm{O(2)} = \rm{SO(2)} \times\rm{B_2}$, 
where $\rm{SO(2)}$ is isomorphic to the group of rotations in 
two dimensions  
and $\rm{B_2}$ 
is the cyclic group of order 2 that is represented by the two matrices 
$1_2$ and ${\rm diag}
(1,-1)$, where $1_2$ is the $2 \times 2$ 
identity matrix. A typical realization of the action of $\rm{SO(2)}$ 
on the variables $\{ y,\beta_2 \}$ is given by
\be
\label{SO2}
    \left( 
	\begin{array}{c}
		\bar{y} \\ \bar{\beta_2}
	\end{array}
    \right) = \left( 
		\begin{array}{cc}
			\cos \th & \sin \th \\
			-\sin \th & \cos \th
		\end{array}
    \right)\;\left( 
	\begin{array}{c}
		{y} \\ {\beta_2}
	 \end{array}
    \right)   .
\ee    
The reflections are given by the product of 
the $\rm{SO(2)}$ rotation matrix and the $\rm{B_2}$ matrix.
By employing Eqs. (\ref{xII}) and (\ref{yII}), together with 
Eqs.  (\ref{beta12})--(\ref{dilaton}), this group 
of transformations may be expressed in terms of 
$\{\alpha,\,\Phi,\,\beta_{\pm}\}$:
\bea
\label{alphaBII} 
\bar{\al} = 
{\frac {8(1+\omega )\left(\cos\theta-1\right)}{33+24\,\omega }} \Phi 
+{\frac{\left(8\,\cos\theta+25+24\,\omega \right)}{33+24\,\omega }} \alpha 
\nonumber \\
-{\frac{4\left(\cos\theta-1\right)}{33+24\,\omega }}\beta_+
+{\frac{4\sin\theta}{\sqrt{33+24\,\omega }}}\beta_-
\\
\label{PhiBII}
\bar{\Phi}= 
{\frac{\left(8 (1+\omega )\cos\theta +3\right)}{11+8\,\omega }} \Phi
+{\frac{8\left(\cos\theta-1\right)}{11+8\,\omega }} \alpha 
\nonumber \\
-{\frac{4\left(\cos\theta-1\right)}{11+8\,\omega }} \beta_+
+{\frac{4\,\sqrt{3}\sin\theta}{\sqrt{11+8\,\omega }}} \beta_- 
\\
\label{beta+BII}
\bar{\beta}_+ = 
{\frac{2\left(1+\,\omega \right)\left(1-\cos\theta\right)}{33+24\,\omega }} 
\Phi
+{\frac{2\left(1-\cos\theta\right)}{33+24\,\omega }} \alpha
\nonumber \\
+{\frac{\left(32+24\,\omega +\cos\theta\right)}{33+24\,\omega }} \beta_+
-{\frac{\sin\theta}{\sqrt{33+24\,\omega }}} \beta_- 
\\
\label{beta-BII}
\bar{\beta}_- = \mp
{\frac{2 (1+\omega ) \sin\theta}{\sqrt{33+24\,\omega }}}\Phi 
\mp{\frac{2\sin\theta}{\sqrt{33+24 \omega }}}\alpha 
\nonumber\\
\pm{\frac{\sin\theta}{\sqrt{33+24 \omega }}} \beta_+ 
\pm\cos\theta\bm. 
\eea
The form of Eqs. (\ref{alphaBII})--(\ref{beta-BII}) implies that 
the symmetry only applies for $\omega > -11/8$. 

The transformations (\ref{alphaBII})--(\ref{beta+BII}) 
are the same for both components of the $\rm{O(2)}$ group. 
The upper and lower signs in 
Eq.  (\ref{beta-BII}) correspond to the 
rotations and reflections of $\rm{O(2)}$, respectively. 
It follows that in 
the case of the $\rm{SO(2)}$ component, for example, 
the scale factors transform to
\bea
\label{a_1BII}
\ln \bar{a}_1 = 
\frac{ 4\,\cos\theta+7+8\,\omega}{11+8\,\omega } \alpha -
\frac{2\,\cos\theta+20+16\,\omega}{11+8\,\omega } \beta_+
\nonumber \\ 
+\frac{2\,\sqrt{3}\sin\theta}{\sqrt{11+8\,\omega }} \beta_- +
\frac{4\left(\cos\theta -1\right)\left(1+\omega \right)}{11+8\,\omega }\Phi 
\\
\label{a_2BII} 
\ln\bar{a}_2 = 
\frac{2\,\cos\theta+2\,\sqrt{11+8\,\omega }\sin\theta+9+8\,\omega}
{11+8\,\omega } \alpha \nonumber \\
-\sqrt{3}\left(\cos\theta-\frac{\sin\theta}{\sqrt{11+8\,\omega }}\right)\bm
-{\frac{\cos\theta-12-8\,\omega 
+\sqrt{11+8\,\omega }\sin\theta}
{11+8\,\omega }} \beta_+ \nonumber \\
+\frac{2\left(1+ \omega \right)\left(\cos\theta-1+\sqrt{11
+8\,\omega }\sin\theta\right)}{11+8\,\omega } \Phi \\
\label{a_3BII} 
\ln\bar{a}_3 = 
\frac{2\,\cos\theta+9
+8\,\omega -2\,\sqrt{11+8\,\omega }\sin\theta}
{11+8\,\omega } \alpha \nonumber \\
+\sqrt{3}\left(\cos\theta+\frac{\sin\theta}{\sqrt{11+8\,\omega }}\right)\bm 
-{\frac{\cos\theta-12-8\,\omega 
-\sqrt{11+8\,\omega }\sin\theta}
{11+8\,\omega }} \beta_+
\nonumber \\
+\frac{2\left(1+\omega \right)\left(\cos\theta
-\sqrt{11+8\,\omega }\sin\theta-1\right)}{11+8\,\omega } \Phi .
\eea
For the reflections,  the transformation of $a_1$ is
given by Eq. (\ref{a_1BII}), but the transformations of ${a_2}$
and $a_3$ differ slightly 
from Eqs. (\ref{a_2BII}) and (\ref{a_3BII}) because they 
depend directly on $\bm$. 

\subsection{Bianchi Types ${\rm {\bf VI}}_{\bf 0}$ and 
${\rm {\bf VII}}_{\bf 0}$}

The elements of $m^{ab}$ for the types ${\rm VI}_0$ 
and ${\rm VII}_0$ are $m^{ab} = {\rm diag} (0, 1, -1)$ 
and $m^{ab} = {\rm diag} (0, 1, 1)$, respectively. 
The curvature potential for each model is given by 
\be
\label{curvature67}
{^{(3)}} R = -2e^{-2\alpha +4\beta_+} f(\beta_- ),
\ee
where
\be
f (\beta_- )=
    \left\{ 
      \begin{array}{cl}
	{\rm cosh}^2 \left( 2 \sqrt{3} \beta_- \right) , 
	     & \qquad {\rm Type} \quad {\rm VI}_0  \\
	{\rm sinh}^2 \left( 2 \sqrt{3} \beta_- \right) , 
	 & \qquad {\rm Type} \quad {\rm VII}_0. 
		\end{array}
    \right.
\ee     
Rewriting Eq. (\ref{curvature67}) in terms of the 
shifted variables (\ref{chi})--(\ref{beta12}) implies that 
\be
{^{(3)}} R = -2 \exp \left[ C_1 \sigma -
\sqrt{\frac{8}{3}} (C_3 \chi -\beta_1 ) \right] f(\beta_2 ),
\ee
where $C_1$ is given by Eq. (\ref{c1}) and  
\be
C_3 \equiv (8+6\omega )^{-1/2}. 
\ee

A discrete symmetry of these models is uncovered by 
defining new variables: 
\bea
\label{x67}
x &\equiv& C_3 \chi -\beta_1  \\
\label{y67}
y &\equiv& \frac{1}{\sqrt{1+C_3^2}} \left( \chi +C_3 \beta_1 \right).
\eea
Action (\ref{bd2}) then takes the form
\be
\label{action67}
S=\int dt e^{-\kappa \sigma} \left[ \left( \frac{8+6\omega}{9+6\omega} 
\right)
\dot{x}^2 +\dot{y}^2 +\dot{\beta}_2^2 -\dot{\sigma}^2 
-2e^{C_1 \sigma -\sqrt{8}x/\sqrt{3}} f(\beta_2)
-2\Lambda \right]
\ee
and Eq. (\ref{action67}) is independent of the variable $y$. This 
implies that the Bianchi types ${\rm VI}_0$ and ${\rm VII}_0$ are 
symmetric under the discrete 
duality transformation
\be
\label{sfd67}
\bar{x}=x, \qquad  \bar{y} =\pm y , \qquad  \bar{\sigma} =\sigma , 
\qquad \bar{\beta}_2 =\pm \beta_2 .
\ee
Substituting these transformations into Eqs. 
(\ref{beta12})--(\ref{dilaton}) implies that 
\bea
\label{baralpha7}
\bar{\al} =  {\frac { 5+6\w}{9+6\w}} \alpha
-{\frac {4(1+\w)}{9+6\w}}\Phi -{\frac {4}{9+6\w}} \beta_+ \\
\label{barphi7}
\bar{\Phi} =  -{\frac {4}{3+2\w}} \alpha
-{\frac {1+2\w}{3+2\w}}\Phi -{\frac {4}{3+2\w}} \beta_+
\\
\label{barplus7}
\bar{\beta}_+ =  -{\frac {2}{9+6\w}} \alpha -
{\frac {2(1+\w)}{9+6 \omega}} \Phi
+{\frac {7+6\w}{9+6\w}} \beta_+
\eea
\be
\label{barminus7}
\bar{\beta}_- = \pm \beta_-
\ee
and the scale factors (\ref{scalefactors}) therefore transform to 
\bea
\label{sfinvert1}
\bar{a}_1  = \exp \left[ \alpha -2 \bp \right]  \\ 
\label{sfinvert2}
\bar{a}_2 =
\exp \left[ \frac{1+2\,\w}{3+2\,\w} \beta_+
+\frac{1+2\,\w}{3+2\,\w} \alpha
-\frac{2\left(1+\w\right)}{3+2\,\w} \Phi \mp 
\sqrt{3} \beta_-\right] \\
\label{sfinvert3}
\bar{a}_3 = \exp \left[ \frac{1+2\,\w}{3+2\,\w} \beta_+
+\frac{1+2\,\w}{3+2\,\w} \alpha 
-\frac{2\left(1+\w\right)}{3+2\,\w} \Phi \pm 
\sqrt{3} \beta_-\right] .
\eea
The scale factor $a_1$ is invariant
under the duality transformation (\ref{sfd67}) 
and $a_2$ and $a_3$  transform non--trivially. 
This is consistent since the curvature potential (\ref{curvature67}) 
depends directly on $a_1^{-2}$ and we have required this term 
to be invariant. 

It is interesting to consider the string cosmologies, where $\omega
=-1$. If $\bar{\beta}_- =-\beta_-$, Eqs. (\ref{scalefactors}), 
(\ref{sfinvert2}) and (\ref{sfinvert3}) 
imply that $\bar{a}_2 = a^{-1}_2$ and $\bar{a}_3
=a^{-1}_3$. Thus, the duality symmetry inverts 
these two scale factors as in the Bianchi type I model. The
presence of curvature implies that the spatial
volume of the universe is not directly inverted, however.  The type I and
${\rm VII}_0$ models are important because they simplify to the
spatially flat FRW cosmology in the isotropic limit.  We conclude that
in the string case, there is a direct generalization of the 
scale factor duality (\ref{sfdfrw}) 
in the type I model, but the symmetry is more restrictive 
in the type ${\rm VII}_0$.

To summarize thus far, we have found that duality symmetries 
exist for the Bianchi types I, II, ${\rm VI}_0$ and ${\rm VII}_0$. We 
will discuss types VIII, IX and V in the next section, 
together with the Kantowski--Sachs and LRS Bianchi type III 
models.

\section{Other Bianchi Types and the 
Kantowski--Sachs Model}

\setcounter{equation}{0}
 
\def\theequation{\thesection.\arabic{equation}}

\subsection{Bianchi Types V, VIII and IX}

The O(3) invariance of the kinetic sector of the Bianchi class A
action (\ref{bd2}) implies that the linear transformations discussed in the
preceding sections are specified in terms of a maximum of three 
parameters. 
Spatial curvature restricts the group of transformations
that leave the full action invariant and therefore reduces the
symmetry. For the type II model, one free parameter remains,
corresponding to an arbitrary angle of rotation.  For types ${\rm
VI}_0$ and ${\rm VII}_0$ the curvature potential 
(\ref{curvature67}) is a sum of 
three independent exponential terms 
and requiring each of these to be invariant
under a linear transformation 
of the variables implies that at least one degree of
freedom must be lost per term. Thus, all three parameters must be
uniquely specified and consequently the action can only be symmetric under
a discrete transformation.

The curvature potentials for the Bianchi types VIII and  IX 
are 
\bea
\label{eight}
{^{(3)}}R_{\rm VIII} =-\frac{1}{2} e^{-2\alpha} \left[ 
e^{-8\beta_+} +4e^{4\beta_+} \left( {\rm cosh} 2 \sqrt{3} \beta_- \right)^2 
+4e^{-2\beta_+} {\rm sinh} 2\sqrt{3} \beta_- \right]
\\
\label{curvature89}
{^{(3)}}R_{\rm IX}
 = -\frac{1}{2}e^{-2\alpha} \left[ e^{-8 \beta_+} +4 e^{4 \beta_+} 
\left( \sinh 2 \sqrt{3} \beta_- \right)^2 
- 4 e^{-2\beta_+} {\rm cosh} 2 \sqrt{3} \beta_- \right] ,
\eea
respectively. 
Both potentials contain 
six separate terms that represent different exponential functions 
of $\{ \alpha , \beta_{\pm} \}$. This 
implies that at least six free parameters are required if the 
cosmological field equations are to be symmetric under linear transformations
of the dynamical variables. 
Since no more than three are available, however, it follows that  
no symmetries of the form 
considered here exist for these two models. 

A similar argument applies to the Taub universe \cite{taub}. 
This is a special case of the type IX model and the action is 
determined by Eqs. (\ref{bd2}) and (\ref{curvature89}) 
with $\beta_- =0$.  
The kinetic sector of the Taub action is invariant under 
a continuous O(2) transformation characterized 
by one free parameter. However, the curvature potential (\ref{curvature89}) 
contains two independent terms when $\beta_- =0$, 
so a minimum of two free parameters are required 
for the full action to be symmetric. We conclude, therefore, that neither 
the Bianchi type IX universe nor its isotropic 
limits (the Taub model and spatially closed FRW 
cosmology) exhibit duality symmetries that result 
in  non--trivial transformations on the scale factors. 

We may also consider the Bianchi type V model (\ref{Vmetric}). 
It follows from
Eq. (\ref{curvature5}) that the contribution of the spatial curvature
to the type V Lagrangian is of the form ${^{(3)}}R e^{3\alpha -\Phi}
\propto e^{\alpha -\Phi}$. Comparison with Eq. (\ref{sigma}) then
implies that the curvature term and the shifted dilaton field cannot
simultaneously remain invariant under non--trivial linear
transformations of $\alpha$ and $\Phi$. This implies that there is no
scale factor duality for this type V model nor its isotropic 
limit (the $k=-1$ FRW universe). These examples highlight 
the inhibitive effect of curvature on the existence of 
dualities.

\subsection{Kantowski--Sachs and LRS Bianchi Type III}

Rewriting the action (\ref{ksaction}) in terms of the shifted variables 
(\ref{chi}) and (\ref{sigma}) implies that
\be
\label{ks1}
S=\int dt~e^{-\ka\sig} \left[ \dot{\chi}^2+\dot{\beta}^2_1-\dot{\sig}^2+
2k\exp \left( C_1 \sigma -\frac{2}{\sqrt{6}} \left( C_4 \chi +\beta_1 
\right) \right) -2\Lambda \right] ,
\ee
where in this subsection 
$\beta_1 \equiv \sqrt{6} \beta$ 
and $C_4 \equiv [2/(4+3\w)]^{1/2}$ is a constant. 
Defining  new 
variables
\bea
\label{u}
u &\equiv&  C_4 \chi+\beta_1 \\
\label{v}
v &\equiv& 
 \left( \frac{4+3\omega}{6+3\omega} \right)^{1/2} ( \chi -C_4 \beta_1 ) 
\eea
implies that action (\ref{ks1}) takes the form
\be
S=\int dt e^{-\ka\sig}\left[ \left( 
 \frac{4+3\w}{6+3\w} \right) \dot{u}^2 + \dot{v}^2
 -\dot{\sig}^2 +2k 
e^{C_1 \sigma - 2u/\sqrt{6}} -2 \Lambda \right].
\ee
This action is cyclic in $v$ and there 
exists a discrete ${\rm Z}_2$ duality:
\be
\bar{u} =u, \qquad \bar{v} =-v , \qquad \bar{\sigma} = \sigma. 
\ee
Eqs. (\ref{average}), (\ref{dilaton}), (\ref{u}) and (\ref{v}) 
then imply that this is equivalent to 
\bea
\label{ksalpha}
  \bar{\al} =  \frac{4+3\omega}{3(2+\omega)} \alpha - 
\frac{2(1 +\omega)}{3(2+\omega )} \Phi +\frac{4}{3(2+\omega )} \beta \\
\label{ksphi}
\bar{\Phi} =  -\frac{2}{2+\omega} \alpha -\frac{\omega}{2+\omega} \Phi 
+\frac{4}{2+\omega} \beta \\
\label{ksbeta}
\bar{\beta} =  \frac{2}{3(2+ \omega )} \alpha +\frac{2 (1+\omega )}{3 (2
+\omega )} \Phi + \frac{2+3\omega}{3 (2+\omega )} \beta 
\eea
in $\{\al,\,\beta,\,\Phi\}$ space 
and Eq. (\ref{kssf}) then implies 
that the corresponding scale factors transform to 
\bea
\label{kssft1}
\bar{a}_1 = \exp\left[\frac{\w \al-2(1+\w)\Phi-2\w \beta }{2+w}\right] \\
\bar{a}_2 = \exp[\al+\beta ] .
\eea
As with  the Bianchi types ${\rm VI}_0$ and ${\rm VII}_0$, 
one of the scale factors is invariant 
under the duality transformation because the curvature potential 
${^{(3)}}R \propto a^{-2}_2$. Furthermore, when $\omega =-1$, 
Eqs. (\ref{kssf}) and (\ref{kssft1}) imply that $a_1$ is inverted, 
$\bar{a}_1 = a^{-1}_1$. 

\section{Discussion and Conclusions}

\setcounter{equation}{0}
 
\def\theequation{\thesection.\arabic{equation}}

In this paper, we have found a number of discrete 
and continuous symmetries of the 
spatially homogeneous, vacuum Brans--Dicke cosmologies containing a
cosmological constant in the gravitational sector of the theory.  
These symmetries relate
inequivalent cosmological solutions via linear transformations on the
configuration space variables $\{ \alpha , \Phi , \beta_+ , \beta_-
\}$.  The dilaton field, $\Phi$, plays a crucial role in the analysis
because it transforms in such a way that the shifted field $\sigma$
may remain invariant even though the cosmological scale factors
transform non--trivially. 

There exists a discrete ${\rm Z}_2$ scale factor duality in the
isotropic $k=0$ FRW model and this symmetry is broken when $k\ne 0$. The
duality is extended to a continuous O(3) symmetry for the spatially
flat, anisotropic Bianchi type I model when $\omega \ne -4/3$. The
symmetry is restricted to O(2) for the type II if $\omega  > 
-11/8 $. There is a discrete ${\rm Z}_2 \times {\rm Z}_2$ 
symmetry in 
the types ${\rm VI}_0$ and ${\rm VII}_0$. There are no symmetries of the
action that lead to non--trivial transformations of the scale factors in
the Bianchi types V, VIII and IX models or the Taub universe. The
Kantowski--Sachs and LRS Bianchi type III models exhibit a ${\rm Z}_2$
duality. These conclusions are summarized in Table 1.
They indicate the extent to which spatial curvature and anisotropy
can have inhibitive effects on the allowed forms 
of duality and this could be of potential significance
for the pre-big bang scenario, given that scale factor duality plays a 
central role in this model. 

\begin{table}
\begin{center}
\begin{tabular}{||c||c|c|c||} \hline \hline

& Negative & Flat & Positive \\
\hline
\hline
& & & \\
Anisotropic   &V         &${\rm {\bf VII}}_{\bf 0}$  &  IX / T  \\
	      &          &               &   \\
Isotropic     & ${\rm F}_{-1}$ & {\bf F}$_{\bf 0}$ &${\rm F}_{+1}$  \\
& & & \\
Anisotropic   &{\bf III}        &{\bf I}&
{\bf KS}  \\
& & &  \\ \hline \hline
\end{tabular}
\end{center}
\vspace{.35in}

\small{Table 1: 
A classification of spatially homogeneous Brans--Dicke 
cosmologies that 
exhibit duality symmetries, where ${\rm F}_k$, KS and T denote 
the FRW, Kantowski--Sachs and Taub universes, respectively. 
The boldfaced models are symmetric at some level. The scale factor 
duality of the flat FRW model can be extended to the anisotropic 
generalizations of this universe. However, neither the spatially curved 
${\rm F}_{\pm 1}$ models nor their direct  anisotropic counterparts
are invariant under duality symmetries. }

\normalsize

\end{table}

The action of the 
continuous O(3) symmetry of the Bianchi type I model corresponds
to rotations on the two--sphere $\chi^2+\beta^2_1 +\beta_2^2
=r^2$, where $r$ is a constant. The spatial
curvature of the Bianchi type II model implies that only the subgroup
of rotations that leave the variable (\ref{xII}) invariant represents
a symmetry of the Lagrangian. This is formally equivalent to restricting
the rotations to the one--surface where the plane $x =
C_2 \chi +\beta_1 = {\rm constant}$
intersects the two--sphere. The intersection corresponds to a circle
and the symmetry is therefore reduced to O(2).
In the Bianchi types ${\rm VI}_0$ and ${\rm VII}_0$, both $|
\beta_2 |$ and the variable $x=C_3 \chi -\beta_1$
must remain invariant. For
example, when $\bar{\beta}_2 =\beta_2$, the action of the full group
of O(3) rotations must be simultaneously restricted to the
two--points on the two--sphere that are intersected by both the
$\beta_2={\rm constant}$ and $x= {\rm constant}$
planes. Thus, the symmetry becomes discrete. 

An analogous interpretation applies 
for the Kantowski--Sachs and LRS
Bianchi type III models. In these cases, 
the configuration space is three--dimensional and spanned by $\{
\alpha , \Phi, \beta \}$. The kinetic sector of the action is 
O(2) invariant under arbitrary rotations on the circle 
$\tilde{u}^2+v^2 =r^2$, where $\tilde{u} \equiv 
\sqrt{(4+3\omega )/
(6+3\omega )} u$ and the variable $u$ is defined in Eq. (\ref{u}). 
When spatial curvature is introduced, 
the symmetry corresponds to the subgroup 
of O(2) transformations that also leave the variable $u= C_4 \chi+\beta_1$
invariant. As a result, the symmetry of the full action is discrete and 
represents a map between the two
points on the circle that are intersected by the line
$\tilde{u}={\rm constant}$. 

The symmetries discussed in this 
paper have a number of applications. 
The symmetric nature of the Bianchi class A 
scalar--tensor cosmologies was recently 
studied in a different context in Ref. 
\cite{lidsey1}. In that paper the existence of point symmetries in the 
cosmological field equations was considered. 
A point symmetry of a set of coupled 
differential equations may be identified by defining a 
vector field  \cite{noether,point1,point}
\be
{\rm {\bf X}} \equiv X_n \frac{\partial}{\partial q_n} + \frac{dX_n}{dt} 
\frac{\partial}{\partial \dot{q}_n} ,
\ee
where  $X_n(q)$ are a set of differentiable 
functions of the configuration 
space variables $q_n$. 
This field belongs to the space that is tangent to the configuration 
space. Contracting the 
Euler--Lagrange equations
\be
\frac{d}{dt} \frac{\partial L}{\partial \dot{q}_n} =\frac{\partial 
L}{\partial q_n} 
\ee
with the functions $X_n(q)$ then implies that 
\be
\label{ld}
\frac{d}{dt} \left( X_n \frac{\partial L}{\partial \dot{q}_n} \right)
=\left( X_n \frac{\partial}{\partial q_n} + \frac{dX_n}{dt} 
\frac{\partial}{\partial \dot{q}_n} \right) L
\ee
and the Lie derivative of the Lagrangian 
$L$ with respect to the vector field ${\rm {\bf X}}$ is 
given by the right hand side of 
Eq. (\ref{ld}). It follows, therefore, that 
there exists a Noether symmetry when this Lie derivative vanishes 
and the corresponding 
conserved quantity is given by 
\be
\label{conserved}
{\cal C} \equiv X_n \frac{\partial L}{\partial \dot{q}_n} .
\ee

When a cosmological constant is present, it can be shown 
that point symmetries only exist in the field equations of the 
Brans--Dicke theory and, moreover, that such symmetries 
arise in types I, II, ${\rm VI}_0$ 
and ${\rm VII}_0$, but {\em not} for the types VIII and IX \cite{lidsey1}. 
It is interesting that the same class A types admit duality symmetries
and this suggests that these dualities may be related 
to the point symmetries. Indeed, 
the discrete scale factor duality (\ref{sfdfrw})
of the spatially flat FRW model 
may be embedded within a continuous point symmetry, 
in the sense that the latter may be employed 
to derive a new set of configuration space 
variables that leave the form of the reduced action invariant 
\cite{lidsey1,sfdnoether}.  
A natural question is whether the dualities we have uncovered 
for the more general Bianchi types are related to point 
symmetries in a similar way. 

It is interesting to consider the behaviour of the first integral 
(\ref{conserved}) under duality transformations. In 
the Bianchi type ${\rm VII}_0$, for example, there exists a point symmetry 
when  $X_{\Phi} = 3 X_{\alpha} = 6X_{\beta_+}={\rm constant}$ and 
$X_{\beta_-} =0$ \cite{lidsey1} and it follows that 
\be
\label{C7}
{\cal C}= 12 X_{\beta_+}
e^{3\alpha -\Phi} \left[ \dot{\alpha} +(1+ \omega ) \dot{\Phi} 
+\dot{\beta}_+ \right] .
\ee
Substitution of Eqs. (\ref{baralpha7})--(\ref{barplus7}) into 
Eq. (\ref{C7}) then implies that 
this quantity changes sign, $\bar{{\cal C}} =-{\cal C}$,  under the discrete 
duality transformation. It is therefore 
invariant if the duality transformation 
is performed simultaneously with a time reversal, $\bar{t}=-t$. 

The duality symmetries we have derived are powerful tools for
generating new, inequivalent cosmological solutions from known
solutions. For example, when the cosmological constant vanishes, any
solution to the vacuum Einstein field equations is also a consistent
solution to the Brans--Dicke field equations, where the dilaton is
constant. Thus, one may begin with a known homogeneous, Ricci--flat
space--time \cite{exact} and apply the relevant duality transformations to
generate a cosmology with a 
dynamical dilaton field. The scale factors will also
transform non--trivially. These exact `dilaton--vacuum' solutions will
allow analytical studies of the pre--big bang scenario to be made in
more general anisotropic settings than those considered previously.
This will provide valuable insight into the region of parameter space
that results in superinflation.

These scale factor dualities of the Bianchi universes also play a
central role in studying the effects of a massless scalar axion field,
$\sigma$, in the matter sector of the Brans--Dicke theory. It 
proves convenient to perform the conformal transformation 
\be
\label{conformal}
\tilde{g}_{\mu\nu} = \Omega^2 g_{\mu\nu} , \qquad \Omega^2 \equiv 
e^{-\phi}
\ee
to the `Einstein' frame, where the dilaton 
field is minimally coupled. Action (\ref{bd}) then takes the form 
\be
\label{conformalaction}
S=\int d^4 x \sqrt{-\tilde{g}} \left[ \tilde{R} -\frac{1}{2} \left( 
\tilde{\nabla} \phi \right)^2 -\frac{1}{2} e^{2\lambda \phi} 
\left( \tilde{\nabla}  \sigma \right)^2 \right]
\ee
where $\phi \equiv (3+2 \omega )^{1/2}$ and $\lambda \equiv [2(3+2 \omega 
)]^{-1/2}$ and we assume that $\Lambda =0$. 
Action (\ref{conformalaction}) is invariant 
under a global SL(2,R) transformation that acts non--linearly 
on the complex scalar field $\chi \equiv \lambda \sigma +ie^{-\lambda 
\phi}$ such that the transformed field is given by $\bar{\chi} 
= (a \chi +b)/(c \chi +d)$, where  
$(a,b,c,d)$ are constants satisfying $ad-bc =1$. The 
Einstein frame metric (\ref{conformal}) remains invariant 
under the transformation\footnote{The reader is referred 
to Refs. \cite{cew,clw} for details.}. This implies that 
the fields 
$\phi$ and $\sigma$ transform non--linearly to
\bea
e^{\lambda \bar{\phi}} = c^2 e^{-\lambda \phi} + (d + c \lambda 
\sigma )^2 e^{\lambda \phi} \\
\lambda \bar{\sigma} e^{\lambda \bar{\phi}} = 
ac e^{-\lambda \phi} + e^{\lambda \phi} (b+a\lambda \sigma)
(d+ c \lambda \sigma )
\eea
and since the dilaton transforms non--trivially,  
the Brans--Dicke metric, $g_{\mu\nu}$, transforms to 
\be
\label{nontriv}
\bar{g}_{\mu\nu} = e^{\bar{\Phi} -\Phi} g_{\mu\nu}
\ee

The importance of this SL(2,R) symmetry is that cosmological 
solutions of the form (\ref{nontriv}) with 
non--trivial $\sigma$ may be derived from solutions where $\sigma$ is 
constant, i.e., from the dilaton--vacuum solutions 
discussed above. Thus, the scale factor dualities of the Bianchi models 
provide the first step in generating anisotropic cosmologies 
with non--trivial dilaton and matter sector from 
vacuum solutions to general relativity. The role of massless  scalar 
fields on the scenario may therefore be studied analytically and possible 
observational effects of 
quantum fluctuations in the matter fields may also be discussed 
along the lines developed in Refs. \cite{cew,clw}. 

Duality symmetries also have important applications in quantum cosmology.
There exists a well known factor ordering problem in the standard
approach due to ambiguities in the quantization of the classical
variables. In the spatialy flat FRW Brans--Dicke model,
however, the scale factor duality (\ref{sfdfrw}) 
may be employed to resolve this
problem \cite{gmv}. 
The natural ordering to choose is the one where the
Wheeler--DeWitt equation remains invariant under the duality
transformation. In principle, the duality symmetries
of the anisotropic cosmologies could be employed in a similar way when
quantizing these models. We also remark that the scale factor duality
of the flat FRW universe is related to a hidden $N=2$ supersymmetry at
the quantum level \cite{lidsey2}.  This leads to a supersymmetric
approach to quantum cosmology that may resolve the problems
encountered in defining a non--negative norm for the wavefunction of
the universe \cite{graham,number}. It would be of 
interest to investigate whether the dualities of the Bianchi 
models are related to hidden supersymmetries. 

Finally, it would be interesting to generalize the analysis to
non--cosmological spacetimes. Cadoni has shown that
$(2+1)$--dimensional Brans--Dicke gravity is O(2) 
invariant when the
metric is static and circularly symmetric \cite{cadoni}. 
The discrete O(2,Z) subgroup of O(2) 
was employed to generate a new spacetime with a
conical singularity that is dual to the black string solution of Ref.
\cite{blackstring}. The question arises as to whether an O(3) symmetry
exists in the four--dimensional Brans--Dicke theory when the metric is 
static. 

\vspace{.7in}

{\bf Acknowledgments} 

\vspace{.1in}

DC and JEL are supported by the Particle Physics and Astronomy 
Research Council (PPARC), UK. 
RT benefited from SERC UK Grant No. H09454.
We thank JWB Hughes and MAH MacCallum for helpful discussions.

\vspace{.7in}



\begin{thebibliography}{99}

\bibitem{scalefactorduality}
Veneziano G 1991 {\em Phys. Lett.} B {\bf 265} 287  \\
Gasperini M and Veneziano G 1991 {\em Phys. Lett.} B {\bf 277} 265

\bibitem{ts}
Tseytlin A A 1991  {\em 
Mod. Phys. Lett.} A {\bf 6} 1721 \\
Tseytlin A A 
and Vafa C 1992 {\em Nucl. Phys.} B {\bf 372} 443 \\
Tseytlin A A 1992 {\em Class. Quantum Grav.} {\bf 9} 979 

\bibitem{gpr}
Giveon A, Porrati M and Rabinovici E 1994 {\em Phys. Rep.} {\bf 244} 77

\bibitem{meissner} 
Meissner K A and Veneziano G 1991 {\em  
Phys. Lett.} B  {\bf 267} 33 \\ 
Meissner K A and Veneziano G 1991 {\em  Mod. Phys. Lett.} A {\bf 6} 3397

\bibitem{pbb} 
Gasperini M, S\'anchez N and Veneziano G 1991 {\em Int. J. Mod. Phys.} 
A {\bf 6} 3853 \\
Gasperini M, S\'anchez N and Veneziano G 1991 
{\em Nucl. Phys.} B {\bf 364} 365 \\ 
Gasperini M and Veneziano G 1993 {\em Astropart. Phys.} {\bf 1} 
317 \\
Gasperini M and Veneziano G 1993 {\em Mod. Phys. Lett.} A {\bf 8} 3701 \\
Gasperini M and Veneziano G 1994 {\em Phys. Rev.} D {\bf 50} 2519 \\
Gasperini M 1995 
{\em Proceedings  of the Second Journ\'ee 
Cosmologie} ed H de Vega and N S\'anchez 
(Singapore: World Scientific) p 429 \\
Levin J J 1995 {\em Phys. Rev.} D {\bf 51} 462 \\
Levin J J 1995 {\em Phys. Rev.} D {\bf 51} 1536

\bibitem{gracefulexit}
Brustein R and Veneziano G 1994 {\em  Phys. Lett.} B  {\bf 329} 429 \\
Copeland E J, Lahiri A and Wands D 1994 {\em Phys. Rev.} D {\bf 50} 4868 \\
Kaloper N, Madden R and Olive K A 1995 {\em  Nucl. Phys.} B
{\bf 452} 677 \\
Kaloper N, Madden R and Olive K A 1996 {\em
Phys. Lett.} B  {\bf 371} 34 \\
Easther R, Maeda K and Wands D 1996 {\em Phys. Rev.} D 
{\bf 53} 4247

\bibitem{gmv}
Gasperini M, Maharana J and Veneziano G 1996 {\em Nucl. Phys.} 
B {\bf 472} 349  \\
Gasperini M and  Veneziano G 1996
{\em Gen. Rel. Grav.} {\bf 28} 1301 \\
Maharana J, Mukherji S and Panada S 1997 {\em Mod. Phys. Lett.} A {\bf 12} 
447 

\bibitem{easther}
Easther R and Maeda K 1996 {\em Phys. Rev. } D {\bf 54} 7252

\bibitem{lidsey}
Lidsey J E 1997 {\em  Phys. Rev.} D {\bf 55}  3303 

\bibitem{frag}
Coley A and Tavakol R 1992 
{\em Gen. Rel. Grav.} {\bf 24} 835 \\
van Elst H and Tavakol R 1994 
{\em Phys. Rev.}  D {\bf 49} 6460 \\
van Elst H,  Dunsby P and Tavakol R 1995 {\em 
Gen. Rel. Grav.} {\bf 27} 171

\bibitem{tw} 
Turner M S and Weinberg E J 1997 Pre--big bang inflation 
requires fine tuning {\em Preprint} FERMILAB--Pub--97/010--A, 
hep--th/9705035

\bibitem{ven97}
Veneziano G 1997 Inhomogeneous pre--big bang string cosmology 
{\em Preprint} CERN--TH/97--42, hep--th/9703150

\bibitem{vencol} 
Buonanno A, Meissner K A, Ungarelli C and Veneziano G 1997 
Classical inhomogeneities in string cosmology, hep--th/9706221

\bibitem{esta} Estabrook F B, Wahlquist H D and Behr C G 1968
{\em J. Maths. Phys.}  {\bf 9} 497


\bibitem{em} Ellis G F R and MacCallum M A H 1969
{\em Commun. Math. Phys.} {\bf 12} 108

\bibitem{rs} Ryan M P and Shepley L C 1975 {\em Homogeneous Relativistic 
Cosmologies} (Princeton, NJ: Princeton University Press). 

\bibitem{mac} MacCallum M A H 1979 {\em General 
Relativity; an Einstein Centenery Survey} ed S W Hawking 
and W Israel  
(Cambridge: Cambridge University Press) p 533

\bibitem{abelian} 
Busher T H 1987 {\em Phys. Lett.} B {\bf 194} 59

\bibitem{dq} 
de la Ossa X C and Quevedo F 1993 {\em Nucl. Phys.} 
B {\bf 403} 377

\bibitem{grv} Gasperini M, Ricci R and Veneziano G 1993 {\em 
Phys. Lett.} B {\bf 319} 438

\bibitem{e}
Elitzer S, Giveon A, Rabinovici E, Schwimmer A  
and Veneziano G 1995 {\em Nucl. Phys.} B {\bf 435} 147

\bibitem{lozano}
Lozano Y 1995 {\em Phys. Lett.} B {\bf 355} 165 \\
Tyurin E 1995 {\em Phys. Lett.} B {\bf 348} 386

\bibitem{trace}
Giveon A and Rocek M 1994 {\em Nucl. Phys.} B {\bf 421} 173

\bibitem{trace1} 
Alvarez E, Alvarez--Gaume and Lozano Y 
1994 {\em Nucl. Phys.} B {\bf 424} 155

\bibitem{79} Klimcik C and Severa P 1995 {\em Phys. Lett.} B 
{\bf 351} 455 \\
Alekseev A Yu, Klimcik C and Tseytlin A A 1996 {\em Nucl. Phys.} B
{\bf 458} 430 \\
Giveon A, Pelc O and Rabinovici E 1996  {\em Nucl. Phys.} 
B {\bf 462} 53

\bibitem{10} Hewson S F and Perry M J 1997 {\em Phys. Lett.} 
B {\bf 391} 316 \\
Balazs L K and Palla L 1997 Quantum equivalence of $\sigma$ 
models related by non--abelian duality transformations, hep--th/9704137

\bibitem{hew} 
Hewson S 1996 {\em Class. Quantum Grav.} {\bf 13} 1739

\bibitem{bd} Brans C and Dicke R H 1961 {\em Phys. Rev.} {\bf 124} 925 

\bibitem{effective}
Fradkin E S and Tseytlin A A 1985 {\em Phys. Lett.} B {\bf 158}  316 \\
Callan C G, Friedan D, Martinec E J  and Perry M J {\em Nucl. Phys.} B 
{\bf 262}   593 \\
Lovelace C 1986 {\em  Nucl. Phys.} B  {\bf 273}  413  \\
Green M B,  Schwarz J H and Witten E 1987 {\em Superstring theory:
Vol. 1} (Cambridge: Cambridge University Press) \\
Casas J A, 
Garcia--Bellido J and Quiros M 1991 {\em Nucl. Phys.} B  {\bf 361} 713

\bibitem{st}
Bergmann  P G 1968  {\em Int. J. Theor. Phys.} {\bf 1} 25 \\ 
Wagoner R V 1970 {\em Phys. Rev.} D {\bf 1}  3209 \\ 
Nordtvedt K 1970 {\em Astrophys. J.} {\bf 161} 1059



\bibitem{higher}
Appelquist T, Chodos A and  Freund P G O 1987 
{\em Modern Kaluza--Klein Theories} (New York: 
Addison--Wesley) \\
Holman R, Kolb E W, Vadas S L and Wang Y 1991 {\em Phys. 
Rev.} D {\bf 43} 995

\bibitem{pn}
Serna A and Alimi J M 1996 {\em Phys. Rev.}  D {\bf 
53} 3087


\bibitem{solar}
Reasenberg R D {\em et al.} 1979 {\em Astrophys. J.}  {\bf 234} 
L219 \\
Will C M 1993 {\em Theory and Experiment in Gravitational 
Physics} (Cambridge: Cambridge University Press)

\bibitem{lidsey2}
Lidsey J E 1995 {\em Phys. Rev.} D {\bf 52} R5407

\bibitem{cadoni}
Cadoni M 1996 {\em Phys. Rev.} D {\bf 54} 7378

\bibitem{wald} Wald R M 1983 {\em  Phys. Rev.} D {\bf 28} 2218 

\bibitem{sneddon} Sneddon G E 1975 {\em J. Phys.} A {\bf 9} 229 \\
MacCallum M A H 1973 {\em Cargese Lectures in Physics} {\bf 6} ed 
E Schatzmann (New York: Gordon and Breach)

\bibitem{noether}
Capozziello S, Marmo G, Rubano C and Scudellaro P 1996  
N\"other symmetries in Bianchi universes {\em Preprint} 
gr-qc/9606050 

\bibitem{ks} 
Kantowski R and Sachs R K 1966  {\em J. Math. Phys.} {\bf 7} 443 

\bibitem{taub} 
Taub A 1951 {\em Ann. Math.} {\bf 53} 472 


\bibitem{lidsey1} 
Lidsey J E 1996  {\em Class. Quantum Grav.} {\bf 13} 2449

\bibitem{point1}
Marmo G, Saletan E J, Simoni A  and   
Vitale B 1985 {\em Dynamical Systems} (New York: Wiley)

\bibitem{point}
Demia\'nski M, de Ritis R,  Marmo G, Platania G, Rubano C,   
Scudellaro P and Stornaiolo C 1991 
{\em Phys. Rev.} D {\bf 44} 3136 \\ 
Capozziello S   and de Ritis R 1993 
{\em Phys. Lett.} A {\bf 177} 1 \\  
Capozziello S and de Ritis R 1994 {\em Class. Quantum Grav.} {\bf 11} 
107 \\
Capozziello S, Demia\'nski M, de Ritis R  
and Rubano C 1995 {\em Phys. Rev.} D {\bf 52} 3288

\bibitem{sfdnoether}
Capozziello S, de Ritis R and Rubano C 
1993 {\em Phys. Lett.} A {\bf 177} 8 

\bibitem{exact} Kramer D, Stephani H, Herlt E and MacCallum M A H 1980
{\em Exact solutions of Einstein's field equations} (Cambridge: 
Cambridge University Press)

\bibitem{cew}
Copeland E J, Easther R and Wands D 1997 Vacuum fluctuations 
in axion--dilaton cosmologies, hep--th/9701082

\bibitem{clw} 
Copeland E J, Lidsey J E and Wands D 1997 S--duality 
invariant perturbations in string cosmology, hep--th/9705050

\bibitem{graham}
Graham R 1991 {\em Phys. Rev. Lett.} {\bf 67} 1381

\bibitem{number}
D'Eath P D, Hawking S W  and Obreg\'on O 1993  {\em Phys. 
Lett.} B {\bf 302} 183 \\
Asano M, Tanimoto M  and Yoshino N 1993 
{\em Phys. Lett.} B {\bf 314} 308 \\
Obreg\'on O, Socorro J  and Ben\'itez J 1993  {\em Phys. Rev.} 
D {\bf 47} 4471 \\
Obreg\'on O, Pullin J and Ryan M P 1993 
{\em Phys. Rev.} D {\bf 48} 5642 \\
Bene J and Graham R 1994 {\em Phys. Rev.} D {\bf 49} 799 \\
Moniz P V 1996 {\em Int. J. 
Mod. Phys.} A {\bf 11} 4321 

\bibitem{blackstring} 
Horne J H and Horowitz G T 1992  {\em Nucl. Phys.} 
B {\bf 368} 444

\end{thebibliography}
\end{document}